\documentclass[prd,superscriptaddress,showpacs,showkeys,amsmath,amssymb]{revtex4}

\usepackage{amssymb,amsfonts,slashed}
\usepackage{graphicx}
\usepackage{epsfig}
\newcommand{\be}{\begin{equation}}
\newcommand{\ee}{\end{equation}}
\newcommand{\bea}{\begin{eqnarray}}
\newcommand{\eea}{\end{eqnarray}}
\newcommand{\beq}{\begin{eqnarray}}
\newcommand{\eeq}{\end{eqnarray}}
\def\({\left(}
\def\){\right)}
\def\[{\left[}
\def\]{\right]}
\def\a{\alpha}

\def\o{\omega}

\def\bp{{\mathbf p}}
\def\bq{{\mathbf q}}

\def\s#1{\slashed{#1}}
\def\t{\tilde}

\def\tr{{\rm tr}}

\begin{document}

\title{Finite temperature Casimir effect for graphene}

\affiliation{Instituto de F\'isica, Universidade de S\~ao Paulo,
Caixa Postal 66318 CEP 05314-970, S\~ao Paulo, S.P., Brazil}

\affiliation{CMCC, Universidade Federal do ABC, Santo Andr\'e, S.P.,
Brazil}

\affiliation{Department of Theoretical Physics, Saint-Petersburg
State University, St.\ Petersburg 198504, Russia}

\author{Ignat V.~Fialkovsky}
\email{ifialk@gmail.com}

\affiliation{Instituto de F\'isica, Universidade de S\~ao Paulo,
Caixa Postal 66318 CEP 05314-970, S\~ao Paulo, S.P., Brazil}

\affiliation{Department of Theoretical Physics, Saint-Petersburg
State University, St.\ Petersburg 198504, Russia}

\author{Valery N.~Marachevsky}
\email{maraval@mail.ru}

\affiliation{Department of Theoretical Physics, Saint-Petersburg
State University, St.\ Petersburg 198504, Russia}

\author{Dmitri V.~Vassilevich}
\email{dvassil@gmail.com}

\affiliation{CMCC, Universidade Federal do ABC, Santo Andr\'e, S.P.,
Brazil}

\affiliation{Department of Theoretical Physics, Saint-Petersburg
State University, St.\ Petersburg 198504, Russia}

\begin{abstract}
We adopt the Dirac model for quasiparticles in graphene and
calculate the finite temperature Casimir interaction between a
suspended graphene layer and a parallel conducting surface. We find
that at high temperature the Casimir interaction in such system is
just one half of that for two ideal conductors separated by the same
distance. In this limit single graphene layer behaves exactly as a
Drude metal. In particular, the contribution of the TE mode is
suppressed, while one of the TM mode saturates the ideal metal
value. Behavior of the Casimir interaction for intermediate
temperatures and separations accessible for an experiment is studied
in some detail. We also find an interesting interplay between two
fundamental constants of graphene physics: the fine structure
constant and the Fermi velocity.
\end{abstract}
\pacs{12.20.Ds, 73.22.-f} \keywords{graphene, Casimir effect}

\maketitle
%%%%%

\section{Introduction}
Nowadays, graphene does not require any long introduction. Its
exceptional properties (see reviews \cite{gra-rev,RMP}) have put
graphene in the focus of current reserach in condensed matter physics
and far beyond.
The principal feature of graphene is the
linear dispersion law $\omega=v_F k$ ($v_F=c/300$ is a Fermi velocity,
$c$ is the speed of light in vacuum) for quasi-particle fermion excitations above graphene
Fermi surface, which is valid for $N=4$ species of fermion
excitations in a graphene layer. This law is valid for the energies up to
a few eV. The dynamics of quasi-particles and their interaction
with electromagnetic field are therefore well described
by the quasi-relativistic Dirac model \cite{DiracModel}. The propagation of electromagnetic
field in the presence of a graphene layer is governed by the effective
action where the fermionic excitations are integrated out.
To the lowest order this action is determined by the standard polarization operator
of $2+1$ dimensional fermions. Such diagrams have been calculated in a number
of papers, see \cite{Appelquist:1986fd,Gusynin,Pyatkovskiy}. One of the most
spectacular predicitions of the Dirac model is the universal absorption
law for light passing through suspended layers of graphene. This law was
experimentally checked with a high precision \cite{Nair}. There are also other experimental confirmations of the Dirac model, most important of which is the observation of the non-conventional quantum Hall effect, \cite{HE}.

In this paper we study the Casimir interaction at non-zero
temperature of a graphene layer with a parallel metal, using the
quasi-relativistic Dirac model for descriptions of quasiparticles in
a graphene. In the framework of this model the problem was already
addressed in \cite{zerot}, where it was shown that at zero
temperature the interaction between graphene and an ideal conductor
is about $2.6\%$ of the interaction between two ideal conductors
separated by the same distance. Earlier, similar calculations were
done \cite{BV,Bordag:2005by,BGKM} for the hydrodynamic model, which
does not reproduce linear spectrum of low-energy excitations in
graphene. More recently, thermal Casimir interaction between
graphene layers was calculated in \cite{Dobson} in a van der Waals
type approach, i.e. basing on the non-retarded Green's functions.
In \cite{Gomez} in the framework of the quasi-relativistic
Dirac model it was found that the Casimir interaction between two
suspended graphene layers becomes strong at high temperature due to
matter plasmonic response in a graphene layer. In another paper,
Ref.\ \cite{1007}, where graphene was modelled by a combination of
Lorentz-type oscillators, practically no temperature dependence of
the Casimir force was found.
%It is important, therefore, to study the finite
%temperature Casimir interaction of graphene in the framework of the
%Dirac model which has been confirmed by several experiments.

 Temperature
dependence of the Casimir interaction attracts much attention due to
the unresolved problem of its asymptotics for metals at large
separations and finite temperatures \cite{Brevik1, Milton, Buenzli,
Bimonte, Pitaevskii, Pitaevskii2}. Two models of dielectric
permittivity, the Drude model and the plasma one, lead to
essentially different Casimir interaction due to their different
infrared frequency dependence \cite{Sernelius, Brevik2, Astrid1,
Mostepanenko}. The free energy of two metals with parallel surfaces
separated by a vacuum slit $a$, for $a\gtrsim \hbar c/(4\pi k_B T)$
is twice as much for the plasma model than for the Drude one
\cite{Sernelius, Brevik2, Brevik3}. Various theoretical models were
used to substantiate arguments in favor of the high temperature
behavior characteristic for a Drude model \cite{Sernelius2,
Pitaevskii3, Dalvit3, Svetovoy}. Predictions of the models also
differ at shorter separations where most experiments are being
performed. However, a discussion of whether one of the models is
excluded by existing experimental data is still under way
\cite{Decca, Dalvit2}. At the same time, we would like to emphasize that the main predictions
of the paper are insensitive to particular model used for description of conductivity of metal
(see discussion in Section IV.A).

The results of the paper are outlined below in brief. In Section
\ref{PolOp} we derive a polarization operator for 2+1 Dirac fermions
at finite temperature and chemical potential. It allows us to
describe in Sect. \ref{Rcoeff} optical properties of graphene in
terms of the components of polarization operator using the modified
Maxwell equations for classical electromagnetic field in the presence of
a suspended graphene layer. Basing on this, we perform  in Section
\ref{FreeEn} a detailed study of the Lifshitz free energy of a
graphene layer interacting with a parallel metal. In Sect. IV.A we derive the
high temperature asymptotics of the free energy of the system, which appears to be the same as
for two conducting surfaces described by the Drude model, i.e. surprisingly strong for a monoatomic graphene layer.
%
% Therefore, at
%large separations (or high temperatures) the Casimir interaction
%appears to be .
%
In Sect. IV.B we investigate  in detail the contribution of non-zero Matsubara terms, and discuss different scaling regimes of the free energy specific to considered graphene-metal system.
%At the end of this Section
%we reveal another
%peculiarity  of the Dirac model -- breakdown of the perturbation
%theory in $\a$ in the high temperature limit.
In Conclusions we outline our main results.

%In particular, a strong suppression of the transverse electric (TE)
%contribution occurs, the transverse magnetic (TM) part of the free
%energy saturates the ideal metal TM part value.
%The mechanism of such a behavior is governed by an interplay
%between the fine structure constant $\a$ and the Fermi velocity
%$v_F$. It can be well understood in terms of TE and TM reflection
%coefficients.

In what follows we adopt $\hbar=c=k_B=1$.

\section{The Dirac model}\label{PolOp}

The dynamics of quasi-particles in graphene is well described within a $2+1$--dimensional Dirac model. All the most important properties of low-energy excitations are indeed readily incorporated in this model: the linearity of the spectrum at the energies below approximately $1$eV, a very small mass gap (if any, but see discussion in
\cite{Appelquist:1986fd,massgap1,Gusynin,Pyatkovskiy}), and the different ``speed of light" inside the graphene layer $v_F\simeq(300)^{-1}$. Moreover, the interaction of the quasi-particles with either quantum or classical external electromagnetic field is straightforwardly described within this model by a standard  ``covariantization''
of the derivative with help of electromagnetic potential $A_\mu$. Thus one deals with the model described by the following classical action (assuming graphene plane lying at $x^3=0$)
\be
    S= -\frac14 \int d^4x\, F^2_{\mu\nu}+
        \int d^3x
        \bar\psi \slashed{D} \psi
%        - e \bar\psi A_m\tilde\gamma^m\psi
    \label{action}
\ee
with
\begin{equation*}
     \slashed{D} = (i\partial_0-\mu-eA_0)\gamma_0
        +v_F[\gamma^1\(i\partial_1-e{A}_1\) +\gamma^2\(
i\partial_2-eA_2\)] -m\,.
\end{equation*}
Since there are $N=4$ species
of fermions in graphene, the gamma matrices are in fact $8\times 8$, being a direct
sum of four $2\times 2$ representations (with two copies of each of the two inequivalent ones), $\gamma_0^2=-(\gamma^{1,2})^2=1$. The Maxwell action is normalized in such a way that
\be
    e^2\equiv 4\pi\alpha =\frac{4\pi}{137}.
    \label{e}
\ee

The properties of the electromagnetic field (both classical, and quantum) in presence of a single graphene layer can be deduced from the partition function
$$
    Z= \int D[A\bar\psi\psi] e^{iS}.
$$
where $S$ is given by (\ref{action}).
Depending on the problem under consideration we can also constrain the electromagnetic
potential $A$ with some additional conditions. In particular, investigating the Casimir interaction between graphene and parallel ideal conductor we will impose the conductor
boundary conditions at $x^3=a$
\begin{equation}
A_0\vert_{x^3=a}=A_1\vert_{x^3=a}=A_2\vert_{x^3=a}=\partial_3A_3\vert_{x^3=a}=0.
\label{condbc}
\end{equation}
thus making the partition function dependent on the distance between graphene and the metal, $Z\equiv Z(a)$.

As the first step of consideration, one shall integrate out fermions thus obtaining
\be
   Z= \int DA e^{-\frac{i}4 \int d^4x\, F^2_{\mu\nu} +iS_{\rm eff}(A)} \label{Za}
\ee
At a somewhat formal level
\begin{equation}
    S_{\rm eff}(A)\equiv-i \ln\det (\slashed{D}). \label{seffdet}
\end{equation}
Since the action (\ref{action}) is quadratic in $\psi$, the expression (\ref{Za}) is
exact, though one has to give a precise meaning to the determinant (\ref{seffdet}).

In the quadratic approximation the effective action $S_{\rm eff}(A)$ discussed above reads
\begin{equation}
S_{\rm eff}(A)= A \ \raisebox{-3.75mm}
    {\psfig{figure=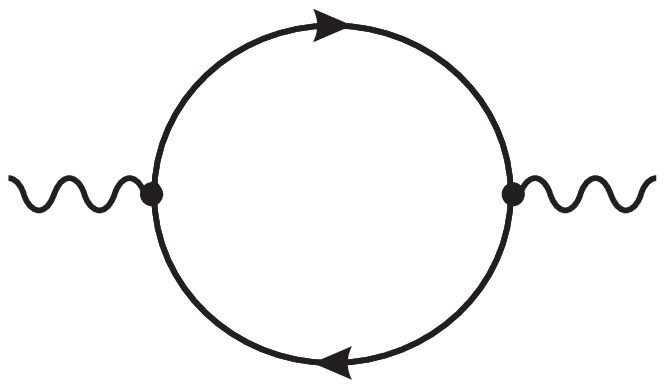,height=.4in}} \  A = \frac 12 \int
    \frac{d^3p}{(2\pi)^3} A_j(p) \Pi^{jl}(p)A_l(p),\label{Seff}
\end{equation}
The polarization operator $\Pi_{ij}$ has been considered in a number of papers for the systems characterized by various sets of parameters. For the zero-temperature case it has been first calculated in \cite{Appelquist:1986fd},
for non-zero $T$ in \cite{DoreyMavratos92}, other cases were considered in, e.g.,  \cite{Pyatkovskiy, LiLiu10}.
Here we present the general result for $\Pi_{ij}(p_0,\bp)$ depending on non-vanishing $m$, $\mu$, $T$. The final formulae (\ref{Pi gen tilda}), (\ref{Pi_00}) coincide with the cases presented in the literature provided one considers appropriate limits of parameters. In particular, (A.27) of \cite{DoreyMavratos92} can be obtained from our result (\ref{Pi_00}) for $m=\mu=0$ and $v_F=1$, while (30) of \cite{LiLiu10} in the case of $\Gamma =0$ is equal to  (\ref{Pi_00}) in the limit $p_0\to0$. The calculations are therefore quite
standard though tedious. We shall be rather sketchy in their description.

In Minkowski space the one-loop polarization operator can be expressed in momentum space as
\be
\Pi^{mn}(p_0,\bp)
    =ie^2 \int\frac{dq_0d^2 \bq}{(2\pi)^3}\,\,
        \tr\( \hat S(q_0,\bq)\tilde\gamma^m  \hat S(q_0-p_0,\bq-\bp)\tilde\gamma^n\),
    \label{Pi_expl}
\ee
where the propagator of the quasiparticles in graphene reads
\be
    \hat S (q_0,\bq) \equiv \slashed{D}^{-1}\vert_{A_\mu=0} =
        -\frac{(q_0+\mu)\gamma_0-v_F\s{\bq}-m}{(q_0+\mu+i\epsilon {\rm\ sgn}q_0)^2-v_F^2\bq^2-m^2}.
    \label{hat S}
\ee
Note that due to the quasi-relativistic nature of excitations in graphene $\hat S$ also depends on the Fermi velocity $v_F $. Here $\bq^j=(q^1,q^2)$, $\slashed{\bq}=\gamma^1q_1+\gamma^2q_2$.

Due to the Lorenz and gauge invariance (transversality) the polarization tensor in an empty space can be completely defined by calculating a single scalar function. In the presence of medium characterized by the velocity $u$ one needs two independent components for complete definition of the (parity-even) $\Pi^{ij}$ \cite{Zeitlin95}
\be
\Pi^{mn}=\frac1{v_F^2}\eta^m_j\[
    \Pi^{ji}_0 A(p_0,\bp)
    +p_0^2 \Pi^{ji}_u  B(p_0,\bp)
    \] \eta_i^n
    \label{Pi gen tilda}
\ee
$$
\Pi^{ji}_0
    =g^{ji}-\frac{\tilde p^j\tilde p^i}{\tilde p^2},\quad
\Pi^{ji}_u
    =\frac{\tilde p^j\tilde p^i}{\tilde p^2}-\frac{\tilde p^j u^i + u^j \tilde p^i}{(\tilde pu)}
    +\frac{u^ju^i}{(\tilde pu)^2}\tilde p^2
$$
where $\eta=\textrm{diag}(1,v_F,v_F)$,
and $A$, $B$ are scalar functions. Note, that in this expression we also took into account the above mentioned quasi-relativistic nature of the excitation in graphene by introducing an appropriate dependence on $v_F$. Vectors with tilde are rescaled by multiplying the spatial components with $v_F$, i.e.,
$ \tilde p^j\equiv\eta^j_i p^i=(p_0,v_F \bp)$. For more details see, e.g., \cite{ChSgraphene}.

In the medium rest reference frame, $u=(1,0,0)$, the
scalar functions $A$, $B$ can be expressed via the temporal
component of the polarization operator and its trace in the
following way ($p_3^2\equiv p_0^2-\bp^2$, $\tilde p_3^2\equiv p_0^2-
v_F^2\bp^2$)
\be
A=\frac{p_3^2}{\bp^2}\Pi_{00}+\Pi_{\rm tr},\quad
    B=\frac1{v_F^2\bp^2} \(\frac{p_3^2+\tilde p_3^2}{\bp^2}\Pi_{00}+\Pi_{\rm tr}\).
    \label{AB}
\ee
Clearly, we can express $A$, $B$ through any pair of the components of $\Pi$. We choose $ \Pi_{\rm tr}\equiv\Pi_m^m$ and $\Pi_{00}$ for convenience of further calculations. In what follows we sketch the derivation of the temporal component $\Pi_{00}$ only, while the trace is calculated similarly.

To introduce the temperature in (\ref{Pi_expl})
 we perform the rotation to the Matsubara frequencies
  (see, e.g. \cite{LandsmanWeert87})
%following the prescription of \cite{LandsmanWeert87}, Eq. (2.3.5)
\be
 i \int dq_0 \rightarrow - 2 \pi T \sum_{k=-\infty}^{\infty}, \qquad
        q_0  \rightarrow 2\pi i T(k+1/2)  ,
        \label{Mats_prescr}
\ee
use the Feynman parametrization
$$
\frac1{ab}=\int_0^1 \frac{dx}{(xa+(1-x)b)^2}
$$
and subsequently change the variables in (\ref{Pi_expl}) in the
spatial part  of the loop--integration:  $\bq\to\bq+x\bp$. Then we
come to
$$
\Pi^{00}=-2 e^2 T N \sum_{k=-\infty}^{\infty}\int_0^1dx
\int\frac{d^2 \bq}{(2\pi)^2}\frac{M_0^2-(q_0+\mu)(p_0-q_0-\mu)}
        {\[(q_0+\mu-xp_0)^2-\Theta^2\]^2}
$$
with $M_0^2=m^2+ v_F^2\bq^2-x(1-x) v_F^2\bp^2$ and
$$
    \Theta^2=m^2+ v_F^2 \bq^2-x(1-x)(p_0^2-v_F^2 \bp^2).
$$

Summation over the fermion Matsubara frequencies can be made explicitly
\be
    \sum_{k=-\infty}^{\infty} \frac1{\[(2 \pi i T(k+1/2)-b)^2-\Theta^2\]^2}=
        -\frac{1}{16 \Theta^3 T^2}
        \(
        \Theta {\rm\ sech}^2\(\frac{\Theta+b}{2T}\)
        -2 T  \tanh\(\frac{\Theta+b}{2T}\)
        \) +(\Theta\to-\Theta).
\ee
The $\bq$--integration can also be performed directly  by using that $\partial_x \tanh x={\rm sech}^2 x $. After these transformations only the integral over $x$
remains, and we arrive at the following representation for $\Pi_{00}$ and
$ \Pi_{{\rm tr}}$
\be
  \Pi_{{\rm tr}, 00}
    =-\frac{2 N \alpha T}{v_F^2} \int_0^1dx \(
         f_{tr,00}\tanh\frac{\Theta_0+b}{2 T}
      -
        \ln\(2\cosh\frac{\Theta_0+b}{2 T}\)
        +(\Theta_0\to-\Theta_0) \)
\label{Pi_00}
\ee
where $\Theta_0\equiv \sqrt{m^2-x(1-x)(p_0^2-v_F^2\bp^2)}$, $b =p_0 x- \mu$, and
\begin{eqnarray}
    f_{00}&=&\frac{-2 v_F^2\bp^2x(1-x) - p_0(1-2x)\Theta_0+
 2\Theta_0^2}{4 T \Theta_0}.
        \label{f_00}\\
    f_{\rm{tr}}&=&\frac{2m^2 v_F^2 + 2 x(1-x) v_F^2  p_3^2}{4 T  \Theta_0}-
 \frac{p_0(1-2v_F^2) (1-2x)- 2(1-v_F^2) \Theta_0}{4T}.
        \label{f_tr}
\end{eqnarray}
We remind that $N$ is the number of fermion species, $N=4$ for graphene. Parity-odd
contributions to the polarization tensor cancel out between different species, while
the parity-even contributions add up.

We note here that the expression (\ref{Pi_expl}) is power-counting divergent.
An ultra-violet divergence indeed appears in the trace part $\Pi_m^m$. To remove this
divergence we performed the Pauli-Villars subtraction at an infinite mass,
which is a rather standard
procedure. One can check that the $m\to\infty$ limit of (\ref{Pi_00}) is indeed zero.

\section{Reflection coefficients}\label{Rcoeff}
In our approximation the full action for the electromagnetic field is
$-\frac{1}4 \int d^4x\, F^2_{\mu\nu}+S_{\rm eff}$, where the effective action (\ref{Seff}) is
confined to the surface of graphene, which we place at $x^3=0$. This
action is singular and gives rise to an interaction
proportional to $\delta(x^3)$ and depending on the tangential momenta.
Therefore, the Maxwell equations receive a singular contribution
\begin{equation}
\partial_\mu F^{\mu\nu} +\delta(x^3) \Pi^{\nu\rho}A_\rho =0.\label{Meq}
\end{equation}
We extended $\Pi$ to a $4\times 4$ matrix with $\Pi^{3\mu}=\Pi^{\mu 3}=0$.
The equations (\ref{Meq}) describe a free propagation of the electromagnetic field
outside the surface $x^3=0$ subject to the matching conditions
\begin{eqnarray}
&&A_\mu \vert_{x^3=+0}=A_\mu\vert_{x^3=-0},\nonumber\\
&&(\partial_3A_\mu)\vert_{x^3=+0}-
(\partial_3A_\mu)\vert_{x^3=-0}=\Pi_\mu^{\ \nu}A_\nu \vert_{x^3=0}
\label{match}
\end{eqnarray}
on that surface. So far, the only difference from the
zero-temperature case \cite{zerot} is in the form of $\Pi_\nu^{\
\mu}$. These matching conditions can also be rewritten in terms of
the electric and magnetic fields $\mathbf{E}$, $\mathbf{H}$.

Next, we introduce a TE mode
\begin{eqnarray}
 && {\bf E}  =(-p_2 {\bf e}_1+p_1 {\bf e}_2) p_0 \Psi(x^3) \\
 && {\bf H}  = i(p_1 {\bf e}_1+p_2 {\bf e}_2)  \Psi'(x^3)
    +{\bf e}_3 (p_1^2+p_2^2)   \Psi(x^3) \\
\end{eqnarray}
and a TM mode
\begin{eqnarray}
&&  {\bf E}  =i(p_1 {\bf e}_1+p_2 {\bf e}_2)   \Phi'(x^3)
    +{\bf e}_3 (p_1^2+p_2^2)  \Phi(x^3) \\
&&  {\bf H}  = (p_2 {\bf e}_1 - p_1 {\bf e}_2) p_0   \Phi(x^3) \\
\end{eqnarray}
where ${\bf e}_1,\ {\bf e}_2, {\bf e}_3$ are unit vectors. An overall factor
of $\exp (i(x^0 p_0+x^1p_1+x^2p_2))$ has been omitted for brevity.

To define the scattering data in the TE and TM sectors we take the
potentials in the form \be
    \Psi(x^3) = \begin{cases}
            e^{ip_3x^3}+r_{\rm TE}e^{-ip_3x^3}, \quad x^3<0 \cr
            t_{\rm TE}e^{ip_3x^3}, \quad x^3>0
            \end{cases},\qquad
    \Phi(x^3) = \begin{cases}
            e^{ip_3x^3}+r_{{\rm TM}}e^{-ip_3x^3}, \quad x^3<0 \cr
            t_{\rm TM}e^{ip_3x^3}, \quad x^3>0
            \end{cases}
\ee

The reflection and transmission coefficients are defined by the
matching conditions. After some algebra we obtain \be
    r_{\rm TE} = \frac{ A}{2i p_3 -  A}, \qquad
        t_{\rm TE} = \frac{2 i p_3}{2i p_3 -  A}
    \label{rTETM-gr}
\ee
$$
    r_{\rm TM} = -\frac{  p_3 (A -v_F^2 \bp^2 B)}
                        {2i \tilde p_3^2 -  p_3 (A-v_F^2 \bp^2 B)}, \qquad
        t_{\rm TM} = \frac{2i \tilde p_3^2}
                        {2i \tilde p_3^2 -  p_3 (A-v_F^2 \bp^2 B)} .
$$
For the later use in the Lifshitz formula (\ref{EL})
 for the Casimir energy, one can also rewrite the reflection
  coefficients in terms of the polarization tensor components
\be r_{\rm TM}=\frac{p_3  \Pi_{00}}{p_3 \Pi_{00}  + 2 i \bp^2},
\qquad r_{\rm TE}= - \frac{ p_3^2 \Pi_{00}+ \bp^2 \Pi_{\rm tr}}
            {p_3^2 \Pi_{00} +  \bp^2 (\Pi_{\rm tr} - 2 i p_3)}.
\label{rTETM-grPi}
\ee

The predictions of (\ref{rTETM-grPi}) for the visible light appear
to be essentially the same as for the idealized ($m=\mu=0$)
zero-temperature case \cite{ChSgraphene} since the corresponding
frequencies are much higher then any other scales in a realistic
graphene sample.

We stress that here we are working with free standing samples of
graphene. Otherwise, the substrate will contribute to the reflection
coefficients.

\section{Free energy}\label{FreeEn}

As early as in 1955 Lifshitz demonstrated \cite{Lifshitz} that the
Casimir interaction between two parallel dielectric slabs can be
expressed in a closed form if their dielectric permittivities are
known at the imaginary frequencies. In a number of later works the
original calculation was generalized and refined \cite{Kats,
Reynaud, Bordag:1995jz, Marachevsky1}. In particular, it was shown
by Kats \cite{Kats} that for any two given parallel plane interfaces
separated by the distance $a$ and described by their reflection
coefficients $r^{(1)}_{\rm TE, \rm TM}$, $r^{(2)}_{\rm TE, \rm TM}$
of the TE and TM electromagnetic modes, the Lifshitz free energy
density reads
\begin{equation}
    {\mathcal F}
    =T\sum_{n=-\infty}^\infty\int\frac{d^2\bp}{8\pi^2} \ln [(1-e^{-2p_\| a}r_{\rm
 \rm TE}^{(1)}r_{\rm \rm TE}^{(2)})
        (1-e^{-2p_\| a}r_{\rm \rm TM}^{(1)}r_{\rm \rm TM}^{(2)})] .
        \label{EL}
\end{equation}
where $p_\|=\sqrt{\o_n^2+\bp^2}$, and $\o_n=2\pi n T$ are the
Matsubara frequencies. The reflection coefficients here are assumed
to be also taken at Euclidian momenta $r=r(\o_n,\bp)$.

Choosing as interacting interfaces a suspended graphene film and a
parallel ideal metal, we shall substitute in the above the
corresponding  reflection coefficients.  For the perfect conductor
one has
\be
    r_{\rm \rm TM}^{(2)}=1,\quad
    r_{\rm \rm TE}^{(2)}=-1.
    \label{rTETM-c}
\ee
The reflection coefficients for graphene at Euclidean momenta can be
found after the substitution $p_0 = i 2 \pi n T = i \omega_n$ into
the formulas (\ref{Pi_00}-\ref{f_tr}),(\ref{rTETM-grPi}). More
precisely, in (\ref{rTETM-grPi}), one should replace $p_3$ by
$i\sqrt{\omega_n^2+ \bp^2}$.

In the following section V.A we consider the high temperature (large
separation) asymptotics of the graphene-ideal metal free energy,
while in subsequent section V.B we study the case $m=\mu=0$ analytically and
discuss numerical results.

%Despite rather idealized nature of the ``ideal metal'' model which we use for description of the second %interface, it will be evident from below that at most separations relevant for an experiment, the results
%
%From what follows it will also be evident that predicted effect is almost independent on particular model used %for description of the conductivity of the metal interface for most separations relevant for a possible %experiment. Thus,

\subsection{High temperature limit}

As discussed in the introduction, the high temperature limit is the
most intriguing regime in the Casimir physics. In this limit at
distances $H\equiv 4\pi a T\gg 1$, the zero frequency Matsubara
terms determine the free energy since other Matsubara terms are
exponentially suppressed at these separations. As we show below, at
zero frequency $r_{\rm TM} \sim 1$, $r_{\rm TE} \sim \alpha N v_F^2
|\bp|$  for small values of the wave vector $|\bp|$. Thus the
coefficient $r_{\rm TM}$ acquires the ideal metal value, and this is
the principal feature of graphene at high temperatures. At the same
time, the TE mode free energy for graphene-metal is suppressed at
separations $a \gtrsim   1/(4 \pi  T)$ by a factor $ \alpha N
v_F^2/(a T) \ll 1$, and due to this suppression the difference in
this regime in the results for the ideal metal and the Drude or the
plasma models of a real metal is too small to be observed in
possible experiments in graphene-metal system. For other
distinctions  of the graphene--metal high-temperature asymptotics
see the next subsection.

%insensitive to the particular model used for description of the
%metal conductivity.

As already mentioned the main contribution in the sum in (\ref{EL}) at high $aT$ is given by the zero frequency Matsubara term
 \be
    {\it \mathcal{F}}_0  = \frac{T^3}{4\pi} \int_0^\infty s ds
\ln\(1+r_{\rm \rm TE} e^{-2 a T s}\)
                \(1- r_{\rm \rm TM} e^{-2 a T s}\)
    \label{EhT}
\ee
where we also performed angular integration and re-scaled the
momentum, $s=|\bp|/T$ as compared with (\ref{EL}).
 Further expansion in powers of $aT$ is determined by the $s\to0$ behavior of the integrand in (\ref{EhT}).
To obtain corresponding asymptotics we first note
 that the polarization tensor components at $\o_n=0$ behave as
\be
\Pi_{00} \mathop{\simeq}_{s\to0}
    \frac{2 N T\a}{v_F^2}\(\ln \(2\cosh\frac{m+\mu}{2 T} \)-\frac{m}{2T} \tanh\frac{m+\mu}{2 T}+(\mu\to-\mu)\)+
\label{Pi00_s}\ee
$$
    +\frac{\a N T^2 s^2}{24 m}
        \(2 \tanh\frac{m+\mu}{2 T} +\frac{m}{2T} \tanh^2\frac{m+\mu}{2 T}-m+(\mu\to-\mu)\)+O(s^4)
$$
\be
\Pi_{\rm tr}-\Pi_{00} \mathop{\simeq}_{s\to0}
    \frac{\a N T^2 v_F^2 s^2}{12 m} \(\tanh\frac{m+\mu}{2 T}+(\mu\to-\mu)\)+O(s^4)
\label{Pitr_s}\ee
From this we readily deduce that
\begin{eqnarray}
&&r_{\rm \rm TE} \mathop{\simeq}_{s\to0}
    -\frac{\a N T v_F^2 s}{6 m} \(\tanh\frac{m+\mu}{2 T}+(\mu\to-\mu)\)+O(s^2),
   \label{k0rTE} \\
&&r_{\rm \rm TM} \mathop{\simeq}_{s\to0}
    1-\frac{s v_F^2/(\a N)}{\ln \(2\cosh\frac{m+\mu}{2 T} \)-\frac{m}{2T} \tanh\frac{m+\mu}{2 T}+(\mu\to-\mu)}+O(s^2). \label{k0rTM}
\end{eqnarray}
where we assumed that $m$ and $\mu$ are of the order of the temperature $T$, which can be relevant for some particular graphene samples and/or experimental setups.
Note that all formulas in this section are also valid for $m,\mu\to0$, since $s\to0$ is essentially the same limit as $T\to\infty$.

As we can see, $r_{\rm \rm TE}$ remains small for  the relevant
values of momenta. Taking the leading term in the expansion of the
logarithm in (\ref{EhT}) and using (\ref{k0rTE}) we obtain the
leading TE contribution at large separations corresponding to the
zero frequency term
\be
    {\mathcal F}_{0\rm TE}
         \simeq  -\frac{\a N T v_F^2 }{192 \pi m a^3} \,
 \Bigl(\tanh\frac{m+\mu}{2T}+(\mu\to-\mu) \Bigr).
            \label{E te 0}
\ee

For the TM mode one cannot use the expansion of $\ln (1-r_{ \rm
TM}e^{-2 aTs})$ as it is divergent near $s=0$, see (\ref{k0rTM}).
Instead one has to use the representation
\be
    {\mathcal F}_{0 \rm TM} =  \frac{T^3}{4\pi} \int_0^\infty s ds \( \ln
 \(1-e^{-2 aTs}\)
    +\ln \(1-\frac{r_{\rm \rm TM}-1}{1-e^{-2 aT s}} e^{-2 aT s}\)\)
\label{repETM}.
\ee
The integral of the first term can be taken explicitly leading
to
\be
   {\mathcal{F}}_{0 \rm TM }^{(0)}=
       -\frac{k_B T\zeta(3)}{16 \pi a^2}
       \equiv {\mathcal F}_{\rm Drude}\vert_{T\to\infty}
        =\frac12 {\mathcal F}_{\rm id}\vert_{T\to\infty},
    \label{E tm 0}
\ee
which coincides exactly with the interaction of two real metals at
high temperature described by the Drude model, or, equivalently, one
half of interaction between two ideal metals at high
temperature, ${\mathcal F}_{\rm id}$. For  readability, in (\ref{E tm 0}) we temporarily restored the physical units. In the rest of (\ref{repETM})
one can use an expansion \be
    {\mathcal F}_{0 \rm TM}^{(1)} = -\frac{T^3}{4\pi} \sum_{l=1}^\infty \frac1l \int_0^\infty s ds
    \(\frac{r_{\rm TM}-1}{1-e^{-2 aT s}}\)^l e^{-2 l aT s}.
\ee
 As the main contribution to the asymptotics at $H\to \infty$
 comes from vicinity of $s=0$ we can expand $r_{\rm TM}-1$
   in a Taylor series around this point,
   to obtain in the highest order
\begin{eqnarray}
     {\mathcal{F}}_{0 \rm TM }^{(1)} &\simeq & -\frac{T^3}{4\pi}
\sum_{l=1}^\infty \frac{\(r_{\rm \rm TM}^{(1)}\)^l}l \int_0^\infty
 s^{l+1}(1+O(s))\(\frac{e^{-2 aT s}}{1-e^{-2 aT s}}\)^l ds\nonumber\\
&=&-\frac{T^3}{4\pi} \sum_{l=1}^\infty \frac1l \frac{\(r_{\rm \rm
TM}^{(1)}\)^l C_l}{(2 a T)^{l+2}} \(1+O\(\frac{1}{2 a T}\)\)
    \simeq -\frac{T^3}{4\pi} \frac{r_{\rm TM}^{(1)} C_1}{(2 a T)^3}=
-\frac{\zeta(3) r_{\rm TM}^{(1)}}{16\pi a^3}
    \label{E tm 1}
\end{eqnarray}
where $ C_l=\int_0^\infty s^{l+1} \(\frac{e^{- s}}{1-e^{- s}}\)^l
ds$ and $r_{\rm \rm TM}^{(1)}\equiv \(\frac{\partial}{\partial
s}r_{\rm \rm TM}\)_{s=0}$ which can be deduced from (\ref{k0rTM}).
% The transition of ${\mathcal F}_{0\rm TM}$
%(\ref{repETM}) to asymptotic (\ref{E tm 0}) occurs at distances $aT
%\sim v_F^2/(\a N)$.
In graphene, where $v_F^2/(\alpha N) \ll 1$, this correction is already small
for considered $H=4\pi aT\gtrsim1$, but in other Dirac systems with larger $v_F$,
the transition of ${\mathcal F}_{0\rm TM}$ (\ref{repETM}) to asymptotic (\ref{E tm 0}) would occur only at much higher distances $aT \gg v_F^2/(\a N)$.
We call attention of the reader that in any case this condition does
not define the transition to the high-temperature regime of the
whole free energy (\ref{EL}) which also includes the nonzero
Matsubara terms.  In the next subsection we investigate in  detail
their contribution and define through a scaling argument the
transition condition for graphene-metal system as $ aT\gg {\a \ln
\a^{-1}}/(2\zeta(3)) \simeq 0.015$.

%acquires in a real graphene its asymptotical value (\ref{E tm 0}) at a much higher %$aT\gtrsim 0.06$.

%We would like to emphasize that for Dirac systems the transition to
%high temperature behavior of the zeroth Matsubara term (\ref{E tm
%0}) takes place at $aT \sim v_F^2/(\a N)$. Assuming the
%expansion parameter is small, $v_F^2/(\alpha N) \ll 1$, the correction (\ref{E tm
%1}) can be neglected for $aT \gg v_F^2/(\a N)$. However, this condition should not be mixed

%On the other hand, as discussed in the following section, the whole expression (\ref{EL}) which also includes %the nonzero Matsubara terms acquires in a real graphene its asymptotical value (\ref{E tm 0}) at a much higher %$aT\gtrsim 0.06$.

Thus we can say, that from (\ref{E te 0}), (\ref{E tm 0}) and
(\ref{E tm 1}), and considerations of the next subsection it follows
that for separations $H\gg 0.19$ the main contribution to the
interaction of a suspended graphene layer with ideal conductor comes
from the TM mode leading contribution (\ref{E tm 0}) and
asymptotically constitutes just one half of the interaction between
two ideal metals at large separations. This is induced by the
specific dependence of the TE,M reflection coefficients on the
momenta $s=|\bp|/T$ for small $s$, see (\ref{k0rTE}),(\ref{k0rTM}).
%
%After investigating the nonzero Matsubara terms in the next subsection we will be able to refine the condition
%
As mentioned earlier, such $r_{\rm TM,E}$ behavior
also makes the predictions for large separations
insensitive to the differences between Drude and plasma models of conductivity of a real metal
in possible experiments. A numerical analysis shows that the difference between graphene-ideal metal and graphene-gold free energies becomes small already at $a \gtrsim 100$nm at $T=300$K (see Fig.\ref{ratio1}) and rapidly decreases with higher separations. Moreover, in this limit the interaction of two suspended graphene samples will acquire exactly the same asymptotical value.

As was shown in \cite{zerot}, in the zero-temperature limit the
Casimir force between graphene and perfectly conducting metal is
about $2.6\%$ of the force between two perfectly conducting metals.
As we see, in the opposite limit, $H\gg 0.19$, the ratio of
graphene - ideal metal free energy to the ideal metal - ideal metal
free energy is strongly enhanced. Such an enhancement takes place
due to the non-perturbative structure of the Lifshitz result in the
limit $v_F=0$.

\subsection{Further analysis and numerical results}\label{NumAn}

Given the specific values of $m$ and $\mu$, the Casimir interaction
can be evaluated by making use of the Lifshitz formula, reflection
coefficients and the polarization operator. As our numerical analysis shows,
increasing the mass gap
diminishes the Casimir interaction (see also a corresponding
discussion in \cite{zerot}), while inclusion of a nonzero chemical
potential $\mu$ enhances the interaction. However, the difference in
the values of the free energy for $m, \mu \sim 0.01$eV, which are
reasonable bounds for these quantities in suspended graphene samples, and the $m=\mu=0$ case is
less than one percent. Therefore in what follows we turn our
attention to the important case $m=\mu=0$ and study it in detail.

It is often desirable to have an accurate analytical approximation
of the exact result at different separations. We present such an
expression for the sum of nonzero Matsubara terms in this section.

To obtain an appropriate analytical expression we first note that at
separations $H\gg v_F$ one can put $v_F=0$ in
any nonzero Matsubara term. It is possible due to the exponential
factor in the Lifshitz formula which effectively restrains the
integration over impulse to $ap_\|\lesssim1$. In this case
contribution of the type of $v_F^2 (ap_\|)^2$ can be neglected
compared to $(ap_0)^2=(2 \pi n aT )^2$ due to the smallness of the
parameter $v_F$.  However, our numerical analysis shows, that this
approximation works at smaller distances as well, see below discussion around (\ref{appt}).
%
%The condition $v_F \ll 4\pi T a \lesssim 1$
%determines medium separations. The condition $4\pi T a \gtrsim 1$ is
%a standard definition of large separations in the Casimir effect. We
%obtain the results for the sum of nonzero Matsubara terms at medium
%and large separations which can be readily used for the comparison
%with experiments.

Secondly, in the finite temperature sum of nonzero Matsubara terms
in the Lifshitz formula one can use the reflection coefficients
taken at zero (!)~temperature. The corrections due to finite temperature are suppressed for nonzero
Matsubara terms, so we neglect them in the leading approximation.
Contrary to this, in zero frequency Matsubara terms one has to take
into account the whole structure of the finite temperature
polarization operator entering $r_{TE,M}$.

Under two mentioned above approximations and the condition $m=\mu=0$
the reflection coefficients of a single graphene layer at zero
temperature have the form:
\begin{align}
r_{\rm TM}^{T=0} &= \frac{\pi \alpha N \sqrt{p_0^2 + p_\|^2} }{ \pi
    \alpha N \sqrt{p_0^2 + p_\|^2} + 8 \sqrt{p_0^2 + v_F^2 p_\|^2} }
    \simeq \frac{\pi \alpha N \sqrt{p_0^2 + p_\|^2} }{ \pi
    \alpha N \sqrt{p_0^2 + p_\|^2} + 8 |p_0|}
    \label{rtm}\\
r_{\rm TE}^{T=0} &=  - \frac{\pi \alpha N \sqrt{p_0^2 + v_F^2
p_\|^2} }{
    \pi \alpha N \sqrt{p_0^2 + v_F^2 p_\|^2} + 8 \sqrt{p_0^2 + p_\|^2} }
    \simeq  - \frac{\pi \alpha N |p_0|}
        {\pi \alpha N |p_0| + 8 \sqrt{p_0^2 + p_\|^2} }
    \label{rte}
\end{align}

\begin{figure}
\centering \includegraphics[width=16cm]{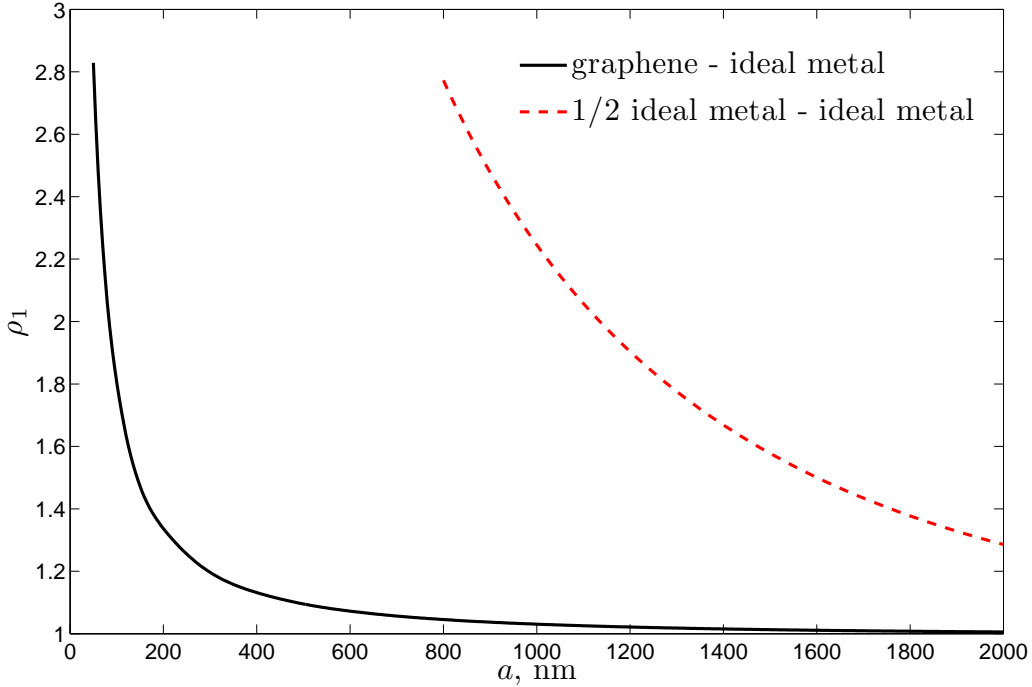} \caption{Ratio
$\rho_1$ of the free energy to the high-temperature asymptotics
$\mathcal{F}_{\rm 0TM}^{(0)}$, see Eq.\ (\ref{E tm 0}).
Both graphs are calculated for $T=300$K. In graphene the values $m=\mu=0$ were
used.} \label{Mayplot}
\end{figure}

Due to smallness of the reflection coefficients (both being of the order
of $\alpha$) we can take just the first term in the expansion of the
logarithm in the Lifshitz formula as another reasonable
approximation for nonzero Matsubara terms. Note, however, that
expansion of the reflection coefficients themselves (at least in TM
mode) is not legitimate, as will become evident below.

\begin{figure}
\centering \includegraphics[width=16cm]{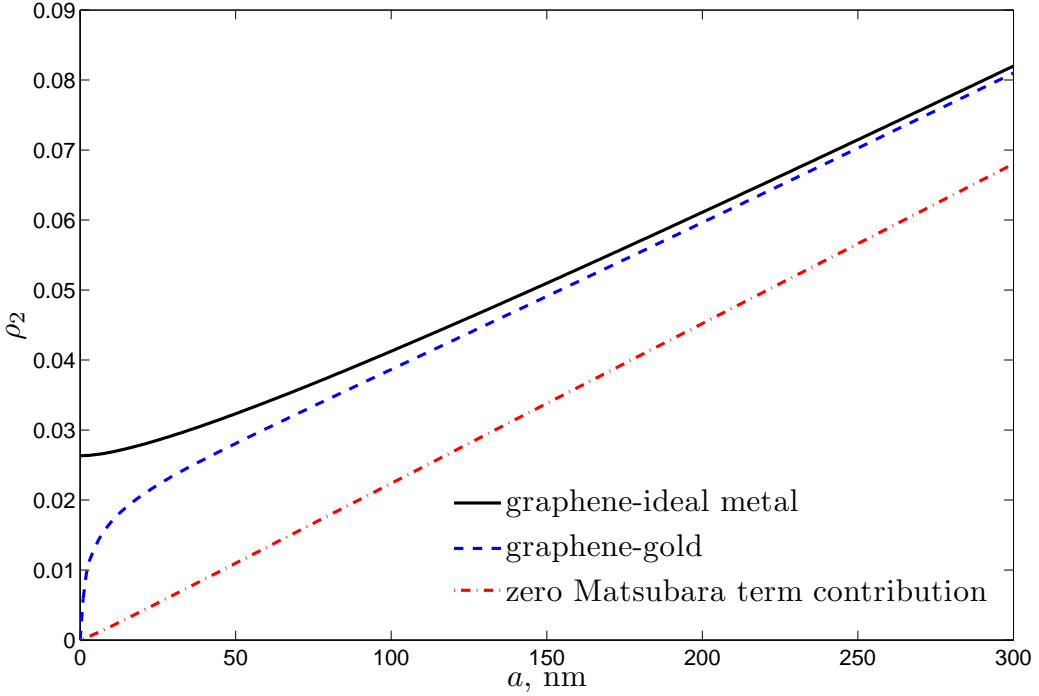} \caption{Ratio
$\rho_2$ of the free energy for a graphene - metal system with
$\mu=m=0$ to the ideal metal - ideal metal free energy at $T=300$K.}
\label{ratio1}
\end{figure}

The sum of nonzero Matsubara terms in this approximation in the TM
case with $r^{(1)}_{\rm TM} = r_{\rm TM}^{T=0}$ (\ref{rtm}) and
$r^{(2)}_{\rm TM}=1$ equals to
\begin{align}
\Delta\mathcal{F}_{\rm TM} &= - \frac{T}{2\pi} \sum_{n=1}^{+\infty}
    \int_{Hn/2}^{+\infty} ds_1 \frac{s_1^2}{s_1+ 16nT/(\alpha N)}
    \exp(-2as_1) = \nonumber \\
    &= - \frac{T}{8\pi a^2} \sum_{n=1}^{+\infty} \exp(-Hn) (1 - gn + Hn)
    + (gn)^2 \exp(gn) {\rm E_1}(gn + Hn) , \label{TMint2}
\end{align}
here $s_1=\sqrt{\omega_n^2+ p_\|^2}$ and $g \equiv 32 T a/(\alpha N)$, $E_1$ stands for the standard
exponential integral function. It is convenient to reexpress the
result (\ref{TMint2}) in an integral form.
For this purpose one has to differentiate
(\ref{TMint2}) over $H$, assuming $H$ as an independent parameter
for the moment, calculate the sum over $n$ and then integrate back
over $H$ (the integration constant is fixed as zero at
$H\to\infty$). Thus one obtains
\begin{equation}
\Delta\mathcal{F}_{\rm TM} = -\frac{T \alpha N}{8 a^2}
    \int_{H}^{+\infty} dt \frac{\exp(t)\bigl(\exp(t)+1 \bigr)\:
    t^2}{\bigl(\exp(t)-1 \bigr)^3 \bigl(8H + \pi\alpha N t \bigr)}
    \label{Intrepr}
\end{equation}
At large separations one can further approximate  $8H + \pi\alpha N
t \approx 8H $ (i.e. expand in powers of $\a$) and get
\begin{equation}
\Delta\mathcal{F}_{\rm TM}\simeq -\frac{\alpha N}{128\pi a^3}
    \Biggl(- \ln(1-\exp(-H)) +\frac{H \exp(-H)}{(1-
    \exp(-H))}  + \frac{H^2 \exp(-H)}{2 (1-\exp(-H))^2 } \Biggr) +O(\a^2)
.\label{TMint3}
\end{equation}
The formula
(\ref{TMint3}) provides an accurate description of (\ref{Intrepr})
for $H \gtrsim 1$ where $\Delta \mathcal{F}_{\rm TM}$ constitutes
less than $6$\% of the whole answer. At $T=300$K this condition
corresponds to distances greater than 600 nm. For smaller
separations it is convenient to use (\ref{Intrepr}) directly.

The TE part of the nonzero Matsubara terms of the Lifshitz formula %can be expanded in $\a$ from the beginning. %
with the coefficients $r^{(1)}_{TE} = r_{TE}^{T=0}$
from (\ref{rte}) and $r^{(2)}_{TE}= -1$   gives  the following
contribution in the leading order in $\a$
\begin{equation}
\Delta\mathcal{F}_{TE} \simeq - \frac{T^2 \pi \alpha N }{8}
    \sum_{n=1}^{+\infty} n \int_{\omega_n}^{+\infty} ds_1 \: \exp(-2 a s_1)
    = - \frac{T^2 \pi \alpha N }{16 a} \frac{\exp(-H)}{(1-\exp(-H))^2} .
    \label{TE1}
\end{equation}

Thus, the complete result for the sum of nonzero Matsubara TM and TE terms in the
approximation described above is given by (\ref{Intrepr}) and  (\ref{TE1})
\begin{equation}
    \Delta\mathcal{F}  = \Delta\mathcal{F}_{TM} +\Delta\mathcal{F}_{TE}.
    \label{nonz}
\end{equation}
Consequently, the leading $v_F=0$ contribution to the free energy is the sum of
(\ref{E tm 0}) and (\ref{nonz}): ${\mathcal{F}}_{0 \rm TM }^{(0)} +
\Delta\mathcal{F}$. It can be used for the comparison of the theory
and experiment almost at all separations.
%Despite formal $4\pi aT\gg v_F$ applicability condition
%
As our numerical analysis shows, a difference between the exact Lifshitz result $\mathcal{F}$ (\ref{EL}) (with $v_F=1/300$, $m=\mu=0$) and considered approximation $ {\mathcal{F}}_{0 \rm TM }^{(0)} + \Delta\mathcal{F}$ is less than $1\%$ for all separations between $1$ and $2000$~nm at the temperature $T=300$K, which approximately corresponds to $3.3 \gtrsim H \gtrsim v_F/2 =1/600$. %(1nm \sim v/2)

In our opinion, the distances corresponding to $H\gtrsim v_F$ (i.e.
above $\sim 2$~nm for $T=300$K) are most relevant for a possible
experiment, and the analytical expressions (\ref{Intrepr}) and
(\ref{TE1}) are obtained to describe this particular regime.
However, numerical studies reveal that $\Delta \mathcal{F}_{TM,E}$
are smooth in the vicinity of $H=0$ point, and collate with $T=0$
result obtained in \cite{zerot}. Indeed, from (\ref{Intrepr}) and
(\ref{TE1}) one gets in this limit
\begin{equation}
\Delta\mathcal{F}\bigl|_{H\to 0}=
    %-\frac{T}{4\pi a^2}\int_{H}^{+\infty} dt \frac{1}{t (t+g)} =
    -\frac{\alpha N}{128\pi a^3} \ln\bigl(1+ 8/(\alpha N \pi)\bigr)
    - \frac{\alpha N}{256 \pi a^3 }. \label{appt}
\end{equation}
where the non analyticity in $\a$ comes from the TM mode, and
explains the precautions we took avoiding to expand the reflection
coefficient in (\ref{rtm}). Since for $T\to0$ the zero-frequency
term $\mathcal{F}_{0}$ (\ref{EhT}) vanishes, eq.~(\ref{appt})
represents the free energy in the leading order in $\a$ in the
approximation when $v_F=0$. The difference between (\ref{appt}) and
the exact Lifshitz result at $T=0$ with $v_F=1/300$ is less than
$1\%$. We also note that at $T=0$ the perturbative expansion in
$\alpha$ is meaningful for $v_F \ne 0$ only, see \cite{zerot}.

%As was anticipated above,  at distances $H\gtrsim 0.75$ the simple
%asymptotical expression $ {\mathcal{F}}_{0 \rm TM }^{(0)}$ (\ref{E
%tm 0}) alone is sufficient for adequate description of the free
%energy.
%As was shown above, in the limit of large separations the Casimir
%free energy has a simple asymptotical expression (\ref{E tm 0}).
%Such behavior is readily verified by the numerical analysis of the exact
%Lifshitz formula.
On Fig.\ref{Mayplot} we plot the ratio $\rho_1$ of the Lifshitz free
energy $\mathcal{F}$ (\ref{EL}) divided by the high temperature
asymptotics ${\mathcal{F}}_{0 \rm TM }^{(0)}$ (\ref{E tm 0}) at
$T=300$K  for two systems: graphene-ideal metal and ideal
metal-ideal metal.  From Fig.\ref{Mayplot} it is evident that the
pace of approaching the asymptotic values by the free energy in the
case  of a graphene -- ideal metal system is much higher
  than in the ideal metal--ideal metal one.
It is explained by the particular behavior of the components of the
polarization operator at nonzero frequencies
$$
    \Pi_{00}(\o_n)\mathop{\simeq}_{|{\bf p}|\to0} \frac{\a {\bf p}^2 c_n}{T} + O({\bf p}^4),\qquad
                \Pi_{\rm tr}(\o_n)\mathop{\simeq}_{|{\bf p}|\to0} \a  T \tilde c_n +O({\bf p}^2),\qquad
                n\geq1 ,
$$
which induce additional $O(\alpha)$ suppression of the corresponding
contributions to the free energy. Here, $c_n$, $\t c_n$ are factors
of the order of unity.  Because of this suppression %of nonzero Matsubara terms
the zero frequency TM Matsubara term dominates in the temperature
dependence of the Lifshitz free energy for a graphene-metal system
at much shorter separations than in the case of a metal-metal
system.

%Above we have discussed various regimes characterizing them by
%smallness of that or other terms in approximate expansions.

It is instructive to define separate regimes on the basis of the
scaling behavior of the free energy, rather than on validity of
various approximations. As a function of distance,
${\mathcal{F}}\sim a^{-3+\delta}$, where $\delta=0$ at $T=0$ and
$\delta=1$ at  $T\to \infty$. The scaling parameter $\delta$ varies
slowly, and can thus be defined through a logarithmic derivative
\begin{equation}
\delta = a \frac{d\ }{da} \big[ \ln a^3 |\mathcal{F}|
\big]\,.\label{delta1}
\end{equation}
Since we are interested in rough estimates, it is enough to
approximate $\mathcal{F}$ by a rational function
\begin{equation}
\mathcal{F}\simeq -\frac{U+aV}{16\pi a^3} \,,\label{Flin}
\end{equation}
where $U$ can be defined by the $T=0$ value (\ref{appt}) and $V$ is
calculated through the high temperature asymptotics, (\ref{E tm 0}).
Namely,
\begin{equation}
 U=\frac {\alpha N}{8} \biggl[ \ln \biggl(1+\frac {8}{\alpha N \pi}\biggr) + \frac 12 \biggr],
\qquad V=T\zeta(3)\,.\label{UV}
\end{equation}
Consequently, (\ref{delta1}) gives
\begin{equation}
 aT=\frac{U\delta}{\zeta(3)(1-\delta)} \,.\label{aTdelta}
\end{equation}
We may consider the ``zero temperature'' regime as the one where $\delta$ is close
to its $T=0$ value, let us say $\delta\lesssim 0.2$, giving $aT\lesssim 0.004$
or $a\lesssim 30{\rm nm}$ at $T=300K$.
%For larger values of $a$ the free energy
%behaves as it should at the ``finite temperature'' region where a linear behavior of $\rho_2$ starts %prevailing, see Fig.\ \ref{ratio1}.
The high temperature regime, on the other hand, may be defined as $\delta\gtrsim 0.8$ corresponding to
$aT\gtrsim 0.06$, or $a\gtrsim 460{\rm nm}$ at $T=300K$. Numerically, this bound is very close to one in the original condition, $H\gg1$, which is universal for any Casimir system. However, in the graphene-metal interaction, %it is of non-asymptotical nature and
as we already discussed, the high-temperature asymptotics is already
saturated at $H\sim1$, see Fig.~\ref{Mayplot}. This can also be seen
by considering an `intermediate' scaling. At distances corresponding
to $\delta=1/2$, where $H\approx 2\pi \alpha \ln
\alpha^{-1}/\zeta(3) \approx 0.19$ is much smaller than unity, the
high temperature asymptotic value already becomes larger than
zero--$T$ contribution, i.e. $aV > U$, and thus this point can be
considered (rather formally) as a crossover between zero and
high-temperature regions.

It is interesting to note that the parameter $U$ which governs
transitions between different regimes is defined by $\alpha$ and
does not depend on $v_F$. One should not however overestimate this
fact. All our expressions are valid for small $v_F$ only. In such a
case the free energy behaves, very roughly, as a sum of the zero
temperature contribution and the zeroth Matsubara term. Therefore,
the crossover between different regimes is defined by the ratio of
these two contributions.

%One can also consider the behavior of the free energy on the shorter separations.
On Fig.\ref{ratio1} we plot the ratio $\rho_2$  of the free energy
of a graphene-metal system to the ideal metal-ideal metal free
energy at separations below $300$~nm. The temperature is chosen as
$T=300$K, which means that $H=4\pi aT$ varies on Fig.\ref{ratio1}
from zero to 0.49. We used a plasma model for gold
$\varepsilon(i\omega) = 1 + \omega_p^2/\omega^2$ with a plasma
frequency $\omega_p = 9.0$eV in numerical calculations of the free
energy for gold semispace and a parallel graphene layer, gold
reflection coefficients are standard Fresnel coefficients for TM, TE
modes with a dielectric permittivity $\varepsilon(\omega)$, see
\cite{Landavshitz}. The numerical result for the free energy of
gold-graphene system divided by an ideal metal - ideal metal free
energy is shown by the dashed blue curve on Fig.\ref{ratio1}. Note
that the behavior of the gold-graphene free energy at short
separations is one power of $a$ different from the graphene-ideal
metal free energy. This is the usual change in power law when one
approaches short separations and the transition from the retarded
regime to a non-retarded one takes place at separations
characterized by a material wavelength $\lambda_p=2\pi/\omega_p$.
Thus, the result for a real metal interacting with graphene is
essentially different from the ideal metal - graphene result at
short separations only. The red curve which corresponds to the zero
Matsubara term only is almost perfectly linear. This reflects the
point that, as mentioned above, the expansion of $\mathcal{F}_{0 \rm
TM}$ around the high temperature asymptotics $\mathcal{F}_{0 \rm
TM}^{(0)}$ is governed by a small parameter $v_F^2/(\alpha N)$,
while the term $\mathcal{F}_{0 \rm TE}$ is suppressed by a factor
$\alpha N v_F^2$.
% Yet again we warn the reader that this condition does not
%define the transition of the whole Lifshitz energy to the high-$T$ regime.

Physical units can be restored in our formulas following the simple rule: in the result expressed as an appropriate power of $aT$ divided by $a^3$, the former product is to be replaced by $k_B a T/(\hbar c)$, while $1/a^3$ is to be substituted by $\hbar c/a^3$.

\section{Conclusions}
We have calculated finite temperature Casimir interaction between graphene and a parallel
conducting plane in the framework of the Dirac model of quasiparticles in graphene.
From the theoretical point of view, it is interesting to note, that we start with
a fully consistent quantum field theory model at finite temperature, and that
the polarization tensor (conductivity of the graphene surface) is thus temperature
dependent.

At high temperature (large separations) the free energy of interaction of suspended graphene sample and a parallel metal asymptotically behaves as the one for two conducting surfaces described by the Drude model, which yields a very strong Casimir interaction, somewhat surprising for a one-atom thick system.
This feature provides an excellent opportunity for the experimental studies of
the temperature Casimir effect in this system at room temperature and large separations as well as for medium ones. The energy at large separations is essentially insensitive to the model which was actually used for description of the metal's conductivity. Moreover, the interaction between two graphene samples would have exactly the same hight temperature asymptotics. Such behavior is induced by the specific static limit ($\omega=0$) of the graphene reflection coefficients, $r_{TM}\simeq1, r_{TE}\simeq0$ given the small Fermi velocity in graphene, $v_F\ll1$.  In this limit they formally coincide with the reflection coefficients for the Drude model of metals.

We studied the Lifshitz free energy for graphene-metal system
in detail and obtained results that can be
readily used for the comparison of the theory and experiment at room
temperature and various separations. For separations corresponding
to $H\equiv 4\pi aT\gtrsim v_F$, we used the limit $v_F=0$ in the Lifshitz formula to
evaluate the sum of nonzero Matsubara terms.
 The speed of quasiparticles in
graphene is much slower than the speed of photons, this is the
physical reason for the accuracy of the $v_F=0$ limit. In this limit
perturbative results in the coupling constant $\alpha$
diverge at $T=0$. Perturbation theory in $\alpha$ does not give
reliable results for graphene systems where $v_F$ is small.

Different physical regimes were separated by investigating the
dependence of the free energy on the distance, $a$. It appeared to
be close to the typical zero temperature one, $a^{-3}$, for $ k_B T
a/(\hbar c)\lesssim 0.004$ and approaching the high energy
asymptotics $a^{-2}$ for $ k_B T a/(\hbar c)\gtrsim 0.06$. The
formal crossover between different regimes can be chosen as $k_B T
a/(\hbar c)\approx0.015$. It is interesting to note that the
metal-graphene system arrives at the high-temperature regime much
faster than the ideal metal-ideal metal system
% In this
%regime the zero frequency TM Matsubara term yields the leading
%contribution to the graphene-metal free energy
due to the $O(\alpha)$ suppression of the TE mode and all nonzero Matsubara terms.
%As a result, graphene yields the same
%asymptotic behavior of the Lifshitz free energy at large
%separations as a metal described by a Drude model.

We have to stress that we have considered a suspended graphene
sample. In such a sample the values of $m$ and $\mu$ are small and,
as our numerical study shows, their influence on the Casimir
interaction is negligible. This approximation will not be applicable
to graphene on a substrate, where one should also take into account
the influence of impurities and the contribution of substrate itself
to the reflection coefficients. Another point of special attention
which should be taken into account when comparing the theory with
experiments is the fact, that exact results in complicated
geometries \cite{Marachevsky1, Marachevsky3,Marachevsky2, Dalvit,
Pirozhenko} can be essentially different from the approximations
based on the Lifshitz result for two parallel plates. In any case,
the state-of-the-art experimental techniques resolve the Casimir
force with total error
of fractions of percent. %(depending on particular configuration).
This permits us believe that a graphene based experiment being not
an easy task is still perfectly feasible.

Finally, a comparison with some previous works on the subject is in order.
%General form of the reflection coefficients for
%a suspended graphene sheet (\ref{rTETM-grPi}) coincides after formal
%transformations with expressions used in \cite{Gomez}, however, the
%expressions for the components of the polarization operators
%entering the reflection coefficients have different forms.
As mentioned before, the leading asymptotics for the free energy of a graphene-metal system
at large separations (\ref{E tm 0}) coincides with one of a graphene-graphene system.
Thus our findings support the results of Ref.\cite{Gomez}
where the thermal van der Waals
interaction in the later system was studied.
However, our prediction for characteristic distance separating zero
and high-temperature regimes (as discussed below
Eq.~(\ref{aTdelta})) differs from that of Ref.~\cite{Gomez}. In our
case it does not depend on the Fermi velocity $v_F$ but does depend
on the fine structure constant $\alpha $, while in~\cite{Gomez} it
is proportional to $v_F$ with no dependence on $\a$.
There is no direct contradiction, however, since different physical systems have been studied in our works.
%Thus a more detailed study of small and medium separations is to be realized in our future work to
%resolve this discrepancy.
%investigate the physical consequences for a graphene-graphene system following from the polarization operator %derived in our paper
%and the one used in \cite{Gomez} requires a separate research at.
%
We disagree with the paper \cite{1007}, which found practically no
temperature dependence of the Casimir interaction of graphene (and
which therefore also contradicts \cite{Gomez}). Still, the estimate
for the zero-temperature case given  in \cite{1007} coincides with
our previous calculations \cite{zerot}. Finally, the separations
considered in \cite{Dobson} are too small to allow for a comparison
with our results.

\begin{acknowledgments}
 This work was supported in parts by FAPESP (I.V.F. and D.V.V.), CNPq (D.V.V.),
 and by the grant RNP 2.1.1/1575 (I.V.F. and V.N.M.).
\end{acknowledgments}

\end{document}